\documentclass[conference]{IEEEtran}
\IEEEoverridecommandlockouts
\usepackage{cite}
\usepackage{amsmath,amssymb,amsfonts}
\usepackage{algorithmic}
\usepackage{graphicx}
\usepackage{textcomp}
\usepackage{xcolor}
\usepackage{booktabs}
\def\BibTeX{{\rm B\kern-.05em{\sc i\kern-.025em b}\kern-.08em
    T\kern-.1667em\lower.7ex\hbox{E}\kern-.125emX}}
\begin{document}

\title{Investigating How MacBook Accessories Evolve across Generations, and Their Potential Environmental, Economical Impacts}

% \author{\IEEEauthorblockN{Zeyi Liao}
% \IEEEauthorblockA{\textit{Depart of Computer Science and Engineering} \\
% \textit{The Ohio State University}\\
% Columnbus, The United States of America \\
% liao.629@osu.edu}

% \and
% \IEEEauthorblockN{Guanqun Song}
% \IEEEauthorblockA{\textit{Depart of Computer Science and Engineering} \\
% \textit{The Ohio State University}\\
% Columnbus, The United States of America \\
% song.2107@osu.edu}

% \and
% \IEEEauthorblockN{Ting Zhu}
% \IEEEauthorblockA{\textit{Depart of Computer Science and Engineering} \\
% \textit{The Ohio State University}\\
% Columnbus, The United States of America \\
% zhu.3445@osu.edu}
% }

\author{
\IEEEauthorblockN{Zeyi Liao}
\IEEEauthorblockA{\textit{The Ohio State University}\\
liao.629@osu.edu}

\and

\IEEEauthorblockN{Guanqun Song}
\IEEEauthorblockA{\textit{The Ohio State University}\\
song.2107@osu.edu}

\and

\IEEEauthorblockN{Ting Zhu}
\IEEEauthorblockA{\textit{The Ohio State University}\\
zhu.3445@osu.edu}
}

\maketitle

\begin{abstract}
    The technological transition of MacBook charging solutions from MagSafe to USB-C, followed by a return to MagSafe 3, encapsulates the dynamic interplay between technological advancement, environmental considerations, and economic factors. This study delves into the broad implications of these charging technology shifts, particularly focusing on the environmental repercussions associated with electronic waste and the economic impacts felt by both manufacturers and consumers. By investigating the lifecycle of these technologies—from development and market introduction through to their eventual obsolescence—this paper underscores the importance of devising strategies that not only foster technological innovation but also prioritize environmental sustainability and economic feasibility. This comprehensive analysis illuminates the crucial factors influencing the evolution of charging technologies and their wider societal and environmental implications, advocating for a balanced approach that ensures technological progress does not compromise ecological health or economic stability.
\end{abstract}


\section{Introduction}
In an era where technological innovation shapes every facet of our lives, from how we communicate to how we work and entertain ourselves, the evolution of charging technologies for consumer electronics, particularly those for the MacBook, offers a compelling lens through which to examine the complex interplay between technological advancement, environmental sustainability, and economic impact. The journey from the introduction of the MagSafe connector to the adoption of USB-C, and subsequently, the return to an updated MagSafe technology, serves as a microcosm of the broader technological landscape, reflecting Apple's responsiveness to shifting technological paradigms, consumer expectations, and the imperative for sustainability.

This comprehensive analysis sets out to unravel the multifaceted implications of these technological shifts, delving into the environmental repercussions associated with the rapid turnover of electronic accessories and the generation of electronic waste. It brings to light the considerable environmental footprint of producing, using, and disposing of charging technologies. Furthermore, this paper explores the economic effects of these transitions on the ecosystem of manufacturers, who are tasked with the continuous innovation and adaptation of production processes, and consumers, who are faced with the financial considerations of upgrading to compatible technologies.

Central to this discussion is the recognition of the critical challenges posed by the burgeoning production of electronic waste, a byproduct of the fast-paced evolution of technology. This issue not only strains waste management and recycling systems but also highlights the urgent need for sustainable manufacturing practices and product life cycle management. Additionally, the paper scrutinizes the economic pressures exerted on both the creators of these technologies, who invest heavily in research and development to stay at the forefront of the industry, and on consumers, who must navigate the cost implications of adopting new technologies.

Through a meticulous exploration of these themes, the paper aims to furnish a nuanced understanding of the dynamics at play in the development of MacBook chargers. It advocates for a balanced approach to technological innovation—one that does not lose sight of the environmental implications and economic realities. By presenting a detailed overview of the evolution of MacBook charging solutions and their wider consequences, the analysis underscores the importance of integrating environmental stewardship and economic pragmatism into the trajectory of technological advancement. This narrative not only enriches our comprehension of the specific case of MacBook chargers but also contributes to the broader discourse on sustainable technological development, challenging us to envision a future where innovation harmonizes with ecological and economic sustainability.

\section{Related Works}
\subsection{Technological Evolution and Consumer Electronics}
To underpin our investigation into the evolution of MacBook chargers, three main themes in related literature are essential. First, studies on the \textit{Technological Evolution and Consumer Electronics} such as those reviewing the progression of charging technologies in vehicles and smartphones, offer valuable analogies and frameworks applicable to MacBook chargers \cite{IEEEexample:acharige2022review, IEEEexample:overview_smartphones}. Second, research on the \textit{Environmental and Economic Impacts} of technology upgrades assesses the lifecycle environmental costs and economic implications, emphasizing the importance of sustainable practices and policies in consumer electronics \cite{IEEEexample:wireless_charging, IEEEexample:magsafe_is_back}. Third, insights into the \textit{Sustainability in Technology Design} from studies on the return of the MagSafe connector and its new capabilities underline the need for integrating environmental stewardship into technological innovation \cite{IEEEexample:magsafe_charger_power}. These areas collectively enhance our understanding of the complex interplay between technological advancement, environmental sustainability, and economic feasibility in the development of MacBook charging solutions.

\subsection{Environmental and Economic Impacts}
Recent studies have illuminated various aspects of the environmental and economic impacts of electronic waste and technological obsolescence. Moheb-Alizadeh and Freire focus on the design of reverse logistics networks to assess the effects of take-back legislation, highlighting the potential for improving e-waste management systems in Washington State \cite{IEEEexample:moheb-alizadeh2023reverse}. In parallel, Koomey and Berard provide a detailed analysis of the carbon footprint associated with Solid State Drives, revealing significant embodied carbon costs \cite{IEEEexample:ssd_embodied_carbon}. On the issue of technological obsolescence, Ma and Wang explore its impacts on firm growth and innovation through patent data analysis \cite{IEEEexample:ma2021technological}, while Kowalkowski discusses strategies for managing obsolescence within digitally transformed SMEs, emphasizing adaptive approaches in product management \cite{IEEEexample:kowalkowski2017managing}. Furthermore, the "Global E-waste Monitor 2024" report adds a broader perspective by quantifying the global impacts of e-waste, underscoring the severe environmental and economic consequences associated with improper electronic waste disposal \cite{IEEEexample:global_ewaste_monitor_2024}.

\subsection{Sustainability in Technology Design}
Recent research explores innovative approaches to enhancing sustainability through technology. Moheb-Alizadeh and Freire utilize machine learning and natural language processing to extract sustainable design insights from online product reviews, highlighting both the opportunities and challenges of implementing such technologies in a sustainable design context \cite{IEEEexample:moheb-alizadeh2023reverse}. Similarly, Schreiber and Meier demonstrate the use of computer vision in deep learning to monitor product wear, aiming to enhance product-service systems and increase sustainability in production and usage \cite{IEEEexample:artificial_intelligence_sustainability}. Khan and Haque provide a comprehensive review of sustainable practices in electronics product design and manufacturing, focusing on material choices, energy efficiency, and end-of-life considerations \cite{IEEEexample:sustainable_electronics_design}. In another vein, Ma and Wang discuss how firms can address the challenges posed by technological obsolescence, particularly how it affects growth, productivity, and capital reallocation in competitive markets \cite{IEEEexample:ma2021technological}. Lastly, the "Global E-waste Monitor 2024" report offers a broad overview of the global impact of e-waste, underscoring the environmental and economic consequences of improper disposal \cite{IEEEexample:global_ewaste_monitor_2024}.

\section{MacBook and Chargers Evlolution}

The MacBook, introduced in 1991, has undergone significant transformations over the years. In this paper, we delve into its evolution, focusing on key milestones and innovations.

\subsection{MacBook Models}

The MacBook line has seen various models, each contributing to its legacy. Notable models include:
\begin{itemize}
    \item Early PowerBooks (1990s): These early laptops laid the foundation for the MacBook series. They featured innovative elements like the setback keyboard and a front-and-center trackball.
    \item iBook G3: A colorful and iconic laptop, the iBook G3 captured attention with its distinctive design and portability.
    \item Latest MacBooks (M-series): Apple’s transition to M-series chips in 2021 marked a significant shift. These new MacBooks combine power efficiency and performance, redefining the user experience.
\end{itemize}

Throughout its evolution, the MacBook introduced groundbreaking features like 1 Trackpad: The PowerBook 500 series introduced the trackpad, revolutionizing navigation and user interaction. 2)Flip-Up Keyboard: The PowerBook G3 featured a unique flip-up keyboard design, enhancing portability and ergonomics.

\paragraph{Designng Change}
Additionally, The MacBook’s design evolved significantly: In 2008, Apple introduced the unibody aluminum design with the MacBook Air. This sleek and lightweight form factor set it apart from its predecessors.

\subsection{Evolution of MacBook Chargers}
In this paper, we speficially focus on the evolution of MacBook chargers, as they are the accessories that are reuqired for each generation regardless of anyother special innovation or not.
\begin{itemize}
    \item MagSafe (2006 – 2011):
    MagSafe was a modular power solution that detached from the computer if the cable was kicked or pulled quickly. Its magnetic connection prevented accidental damage.
    \item USB-C Charging (2009 – Current):
    USB-C chargers became standard, offering versatility across different MacBook models. However, charging speed varied based on the wattage rating of the charger.
    \item MagSafe 2 (2012 – 2017):
    The redesigned MagSafe port accommodated slimmer laptops while maintaining the convenience of magnetic attachment.
    \item Return to MagSafe: Despite the widespread adoption of USB-C, Apple reintroduced a compatible version of MagSafe (MagSafe 3). This decision reflects Apple’s commitment to user experience and adaptability.

\end{itemize}
In summary, Apple’s continuous innovation and responsiveness to user needs drive the evolution of MacBook technology. As we look ahead, we anticipate further advancements that will shape the future of portable computing. However, this is not a sure thing about the benefit of designing new technogies and we will discuss in the later sections.

\section{Experiments}
To more qualitively study the evoluation of MacBook chargers, we collect data from Walmart for each type of charger\footnote{We have tried to collect data from Amazon, however, due to the anti-crawl policied and internal defense mechanisim, we decide to focus on Walmart which is also a popular website for people to purchase chargers in daily life} from 2016 to 2024 and analysis the purchase time by dividing them into two halfs per year.

Specifically, we utilize the API provided by SerpAPI\footnote{https://serpapi.com} which offers feasible and fast API to different search engines including Walmart search engines. Since searching engine only provide approximate search results, we mainly use the key-words L-Tip magsafe, magsafe 2, magsafe 3 and UCB-C to UCB-C for different items respectively and we manully check if the returned results have significant relevance for the searching items.

\begin{figure}[!htbp]
    \centering
    \includegraphics[width=1.1\columnwidth]{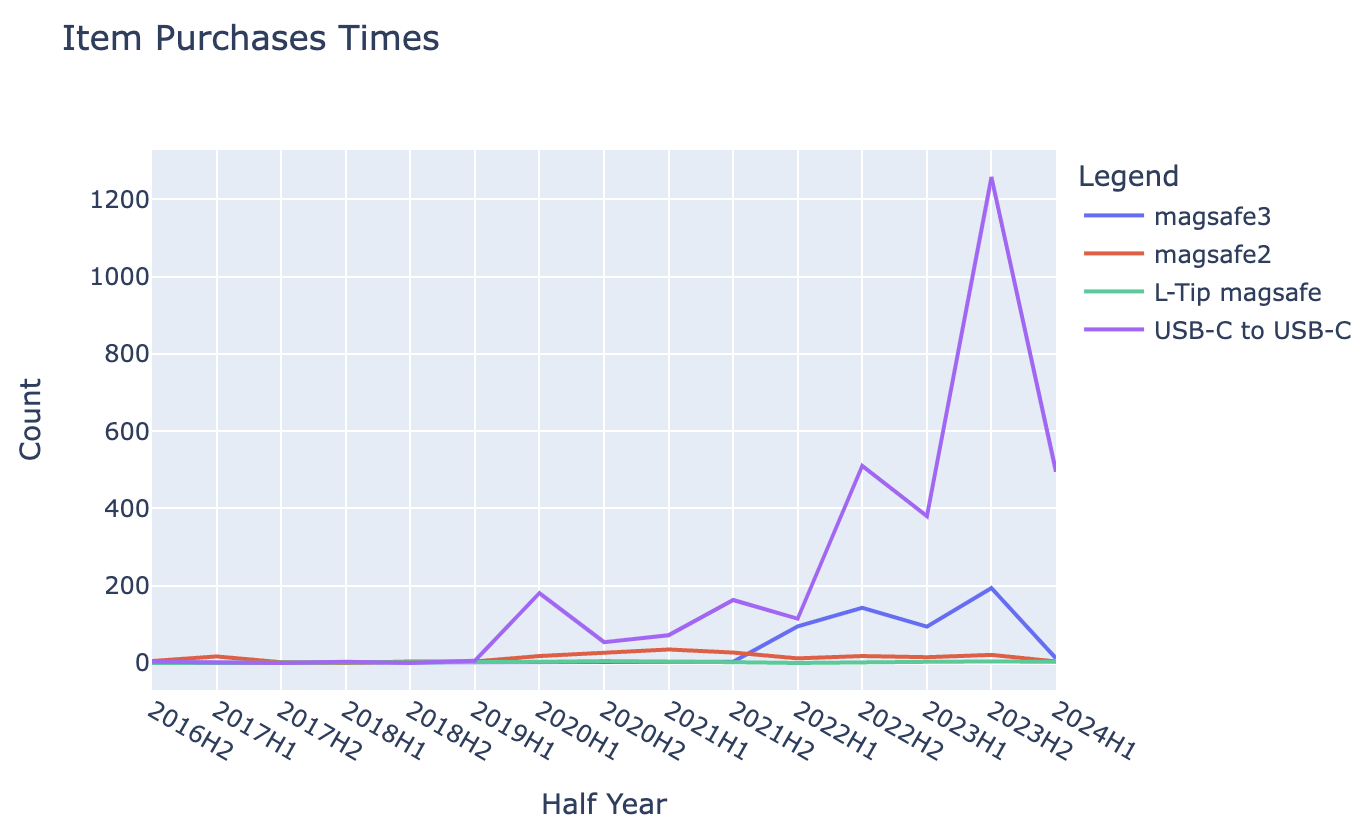}
    \caption{Figures about the purchase times for different chargers. We could observe that Magsafe 1 (L-Tip magsafe in the legend) and Magsafe 2 dont have too many purchase times in the recent years, while USB-C (USB-C to USB-C in the legend) have relatively more times compared to them. We hypothesize this is bacause magsafe 3 charger become popular with newer version of macbooks which are not compatiable to obsoleted chargers like magsafe 1,2. For USB-C charger, it has significant larger purchase amounts compared to others due to it's compatiable to many other devices like phone or Microsoft Computers as well.}
    \label{fig:charger_trend}
\end{figure}

To more accurately capture the variety in charger usage, we plan to analyze user behaviors across the top 10 pages of search results, if available, aggregating the purchase frequency for each item. It's important to note that Walmart's pages may contain advertisements promoting products unrelated to our search criteria. To minimize the inclusion of such irrelevant data—a potential source of false positives—we will focus exclusively on items accompanied by user reviews. This approach ensures that our dataset reflects genuine purchasing activities, providing a more reliable foundation for understanding consumer preferences and trends in charger usage.

\paragraph{Results}
Figures are shown in the Figure \ref{fig:charger_trend}. Obviously, we observe that magsafe 1 and magsafe 2 have nearly no purchase records in the recent years while magsafe 3 and USB-C charger have reletively larger purchase time compared to them. Especially,  the time of purchasement of USB-C charger reach to 1200 times at the second half in 2023. We hypothese that this is because that Apple company initially design the magsafe 1 (L-Tip magsafe) and magsafe 2 for their products, but the products are quickly being obsoleted after iterations of new products in previous years. Given that, the purchasement is sporadic in the recent years with the potential that some people are still sticking to old devices. For magsafe3, it's becuase it's the newest type of mag-series chargers which are required by the latest several version of MacBook. Due to the advanced MacBook's high performance and high efficiency, users who take advantage of them have to use the latest magsafe 3 version which increase the purchasement purchase time over the recent years. Interesting, we notice that USB-C would achieve really high purchase time compared to others. The reason for such increase comes from two lens: 1) USB-C is always compatiable to different version of macbooks no matter being compatiable to magsafe or not. Users could buy UCB-C to ensure their adaptability to different version of macbook at any time. 2)USB-C is not only the charger for MacBook but also could be adapted to different devices like iPhone ors Microsoft laptop, hence, users intend to buy USB-C for their universality. We assume it would furhter increase as time goes by.

\begin{table}[!htbp]\centering
    \caption{Average price for each charger}\label{tab:price}
    \scriptsize
    \begin{tabular}{lcccc}
    \toprule
    & Magsafe 1 & Magsafe 2 & Magsafe 3 & USB-C \\
    \midrule
    Price & 19.22 & 30.976 & 60.02 & 12.01 \\
    \bottomrule
    \end{tabular}
\end{table}

We further include the average price for each type of charger in Table \ref{tab:price}. Although magsafe 1 and 2 only possess very low price compared to magsafe3, they are obsoleted because the older version of mac is knocked out of the game. So there is not direct correlation between the price and the purchase time if the item like charger are required components in the daily life. USB-C is cheaper than magsafe 3 and the high pruchase time from USB-C could partially attributed to the low price of USB-C as well.

However, we highlight the inherent limitations associated with exclusively analyzing online Walmart data. Firstly, it is important to recognize that consumers may opt for offline purchases, which are not captured in our online dataset. Secondly, Walmart is not the sole retailer for these products; consumers also patronize other retailers, both online and offline, which diversifies purchase behaviors beyond our dataset. Thirdly, the recent surge in the popularity of online shopping may skew the data, rendering it less representative of consumer behavior in earlier periods when online purchasing was less prevalent. These factors collectively suggest that while our analysis offers valuable insights, it may not fully encapsulate the breadth of consumer purchasing patterns and preferences.

\section{Why MacBook Change Their Charger}

The evolution of MacBook chargers, specifically the transition from MagSafe to USB-C and then the introduction of MagSafe 3, illustrates Apple's response to technological advancements and regulatory pressures, alongside a commitment to component reusability and environmental sustainability.

\paragraph{Transition from MagSafe to USB-C}
MagSafe, a proprietary magnetic charging connector, was favored for its safety feature, which allowed the cable to detach easily if accidentally pulled, thus preventing the laptop from falling. However, in 2016, Apple shifted to USB-C for several compelling reasons 1) Standardization and Versatility: USB-C has become an industry-standard connector, supporting data transfer, video output, and charging. Its adoption promotes a unified charging ecosystem, simplifying the user experience across various devices.
2) Design Innovations: The slim and compact nature of USB-C ports enabled sleeker laptop designs.
3) Enhanced Charging Capabilities: USB-C supports higher power delivery, facilitating faster charging times for a broader range of devices.
\paragraph{Introduction of MagSafe 3}
Despite the advantages of USB-C, Apple reintroduced MagSafe in 2021 with MagSafe 3, blending magnetic convenience with the efficiency and versatility of USB-C. MagSafe 3, featured in the 14-inch and 16-inch MacBook Pro models, and later the MacBook Air, offers 1) Faster Charging: Leveraging USB Power Delivery (USB PD) 3.0 for optimal power delivery while maintaining compatibility with older chargers.
2) Efficiency and Safety: Engineered for safer and more efficient charging.
3) Backward Compatibility: Ensuring older chargers can still power the latest models, albeit not at full capacity.

\paragraph{Component Reusability and Environmental Considerations}
A deeper dive into the components of MacBook chargers reveals a focus on sustainability and compatibility 1) Switching Power Supplies: Utilizing efficient conversion mechanisms, these power supplies are designed for longevity and cross-generation use.
2) Heat Sinks and Circuit Boards: These essential components are engineered to maximize cooling and efficiency, potentially allowing for their reuse across different charger generations.
3) Innovative Solutions like the AnyWatt MS Adapter: This adapter exemplifies component reusability by enabling MagSafe and MagSafe 2 chargers to power USB-C devices, thus extending their utility and reducing electronic waste.

All in all, the evolution of MacBook charging solutions from MagSafe to USB-C and the reintroduction of MagSafe with MagSafe 3 reflect Apple's strategic adaptation to technological trends, regulatory requirements, and user needs. The emphasis on component reusability across generations not only underscores Apple's commitment to sustainability but also aligns with broader industry trends towards standardization and environmental responsibility. Through these developments, Apple continues to balance innovation with environmental stewardship, ensuring that its products remain at the forefront of technology while contributing to a more sustainable future.

\section{Environmental Impact of Obsoleted Chargers}
The rapid evolution of MacBook chargers from MagSafe to USB-C and back to MagSafe 3 reflects a pursuit of technological innovation and improved user experience. However, this progress comes with environmental costs. The phase-out of older charger models contributes significantly to the accumulation of electronic waste (e-waste), raising concerns about the sustainability of continuous technological upgrades.

\paragraph{E-Waste Accumulation}

Obsolete MacBook chargers add to the growing problem of e-waste, challenging both waste management and recycling efforts. Despite design intentions for recyclability and minimal use of harmful substances, the disposal of outdated chargers contributes to environmental degradation.

\paragraph{Resource Consumption and Emissions}

The life cycle of MacBook chargers, encompassing manufacturing, usage, and disposal, demands significant resources and energy, leading to greenhouse gas emissions. Even as new charger technologies promise improved efficiency, the environmental implications of their production and eventual obsolescence cannot be overlooked, highlighting the need for sustainable innovation.

\paragraph{Innovation vs. Environmental Impact}

Technological advancements in charging solutions, while aiming for better performance and convenience, come with the dilemma of environmental impact. The benefits of new technologies must be weighed against their ecological footprint, emphasizing the importance of developing sustainable practices that align with environmental conservation goals.

All together, as Apple continues to innovate in the realm of MacBook chargers, the environmental implications of such advancements warrant careful consideration. The balance between embracing new technologies and mitigating their environmental impact is crucial. Efforts to reduce e-waste, alongside sustainable design and recycling initiatives, are essential in ensuring that technological progress does not come at the expense of the planet's health.

\section{Economical Impact of Obsoleted Chargers}

The shift towards new charging standards, resulting in the obsolescence of previous charger models, has significant economic implications for both manufacturers and consumers, alongside the costs associated with research and development.

\paragraph{From the Producer's Perspective}

\begin{figure}[!htbp]
    \centering
    \includegraphics[width=1.1\columnwidth]{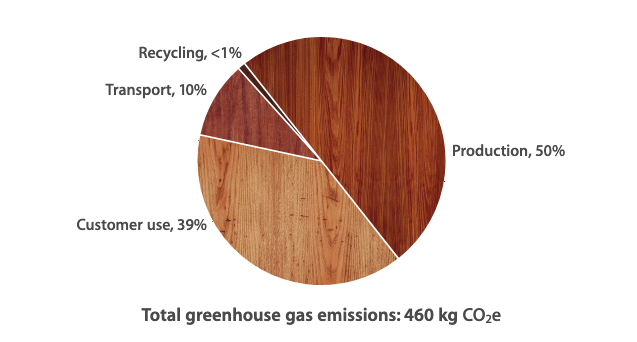}
    \caption{Distribution of Greenhouse Gas Emissions by Life Cycle Stage: The chart illustrates the total greenhouse gas emissions associated with a product, amounting to 460 kg CO2e. Production is the largest contributor, accounting for 50\% of emissions, followed by customer use at 39\%. Transport contributes 10\%, while recycling accounts for less than 1\%, indicating significant opportunities for reducing the environmental impact through enhancements in production processes, increased product-use efficiency, and improved recycling initiatives.}
    \label{fig:apple_gas}
\end{figure}

\begin{figure}[!htbp]
    \centering
    \includegraphics[width=1.1\columnwidth]{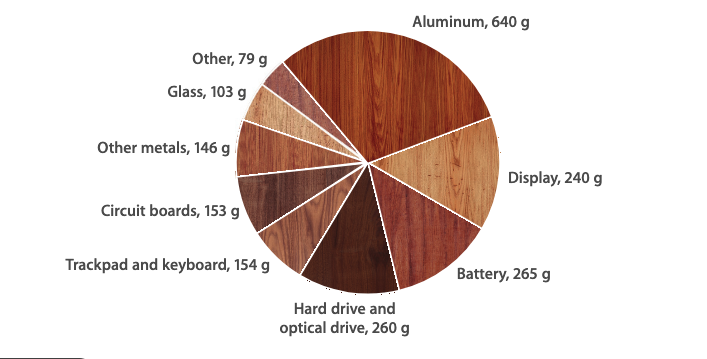}
    \caption{Distribution of Material Weights in a MacBook Laptop. The chart illustrates the proportionate weight of various materials and components that constitute the laptop, highlighting the predominance of aluminum and the significant contribution of the battery to the overall mass.}
    \label{fig:apple_material}
\end{figure}

Manufacturers incur substantial costs in developing new charging technologies, covering research, design, and integration with devices. This investment is a gamble, as the success of new chargers is not guaranteed. Additionally, updating production lines to accommodate new standards can be expensive, potentially affecting the pricing of new devices. We have two figures presenting the exact gas emission and material usage percentage distribution for Apple in Figure 2 and Figure 3.

Particularly, we also conduct a comprehensive analysis for Figure 2 here.
The figure 2 presents a pie chart that illustrates the distribution of total greenhouse gas emissions associated with a product, measured in kilograms of carbon dioxide equivalent (kg CO2e), a standard unit for quantifying the carbon footprint. The chart indicates that the total emissions amount to 460 kg CO2e and are divided into four main categories:
\begin{itemize}
    \item Production (50\%): This is the largest contributor to the product's carbon footprint, accounting for half of the total greenhouse gas emissions. It suggests that the manufacturing processes, from the extraction of raw materials to the final assembly of the product, are energy-intensive and release a significant amount of carbon dioxide and equivalent gases.
    \item Customer Use (39\%): The second-largest share of emissions comes from the customer's usage of the product. This high percentage indicates that the product, once in the hands of consumers, continues to have a considerable environmental impact, possibly due to energy consumption during its operational phase.
    \item Transport (10\%): The transport segment, which encompasses the distribution, shipping, and logistics from the manufacturer to the consumer, represents a smaller but still notable portion of the emissions. This reflects the environmental impact of fuel consumption and exhaust gases emitted by transportation vehicles.
    \item Recycling (<1\%): The smallest slice of the chart relates to the end-of-life phase, specifically recycling. Its minimal contribution to total emissions implies that recycling processes for this product are relatively efficient or that the volume of the product being recycled is currently low.
\end{itemize}

The analysis suggests that efforts to reduce the product's overall carbon footprint could be most effectively focused on the production and customer use phases, which together account for 89\% of total emissions. Strategies to optimize manufacturing efficiency, switch to renewable energy sources, or improve product energy efficiency during use could significantly reduce the environmental impact. Furthermore, the relatively small impact of recycling points to an opportunity to enhance the product's design for recyclability, encouraging a circular economy that could further minimize emissions. However, the relatively higher emissions from transport indicate that logistics optimization and the use of greener transport options could also contribute to overall emission reductions.

Specifically, we also conduct a comprehensive analysis for Figure 3 here.
This figure is a pie chart representing the distribution of materials by weight in a laptop. The chart is designed to showcase the different components and their respective weights in grams. Let's analyze each segment:
\begin{itemize}
    \item Aluminum, 640 g: This is the largest segment of the chart, indicating that aluminum is the heaviest single material used in the construction of the laptop. Aluminum is often used for the laptop's casing due to its light weight and durability.
    \item Battery, 265 g: The second-largest segment is the battery, which is a significant part of the laptop's weight. The battery is crucial for providing power to the laptop, allowing it to operate without being plugged into an electrical outlet.
    \item Hard drive and optical drive, 260 g: Combined, these two components are slightly less in weight than the battery. The hard drive is used for data storage, while the optical drive, which might be less common in modern laptops, is used to read and write CDs or DVDs.
    \item Display, 240 g: The display includes the screen and possibly the front glass or plastic cover. This weight suggests a moderately sized display, as larger or higher-quality displays tend to weigh more.
    \item Trackpad and keyboard, 154 g: These input devices are essential for user interaction with the laptop. The weight suggests they are likely integrated into the main body of the laptop, which is common in modern designs.
    \item Circuit boards, 153 g: This includes the motherboard and other smaller boards. The motherboard is the central printed circuit board (PCB) in many modern computers and holds many of the crucial electronic components of the system.
    \item Other metals, 146 g: This segment covers metals other than aluminum, possibly including steel or copper, which are used in various components for structural support, electrical conduction, and heat dissipation.
    \item Glass, 103 g: This likely refers to the glass used in the display, which may be part of the screen's construction, possibly for touch screen functionality or as a protective layer.
    \item Other, 79 g: This segment is likely to encompass various other materials such as plastics, rubbers, and composites used throughout the laptop for insulation, spacing, and protective purposes.
\end{itemize}

The Figure 3 pie chart suggests that the laptop is primarily composed of metal (aluminum and other metals), which is common for laptops designed to be sturdy and have a premium feel. The relatively high weight of the battery indicates that the laptop is likely designed for extended use between charges, which might be a selling point for users needing portability and long battery life.

\paragraph{From the Consumer's Perspective}

For consumers, the need to purchase new chargers for compatibility with newer devices can lead to extra expenses. Despite potential benefits like faster charging, the immediate financial burden of acquiring new technology is palpable.

\paragraph{Considering the Costs of Innovation}

The economic rationale behind new charger technologies involves a trade-off between the initial high costs of research and development and the anticipated long-term benefits, such as efficiency gains and reduced environmental impacts. However, it's uncertain whether these benefits will outweigh the substantial upfront investment required for innovation.

Put all in together, the transition to new charging technologies presents economic challenges for both producers, who face production costs, and consumers, who may need to invest in new accessories. Balancing the initial investment against potential future savings and benefits is crucial, underscoring the complex economic dynamics of technological advancement in charging solutions.

\section{Issues I Encountered and Contribution}
During collect data about user behaviors, I first try to crawl data from Amazon which is the most popular online shopping website. However, because of the strong defense mechanisim and restricted policies, I tried many ways and many online repo to do it but failed in the end. So I choose the second popular online website Walmart.

Although there is an API to save my life, but the search results are not accurate enough like the results sometimes differ from my searching items. Based on that, I just mannully check each searched results and ensure their relevance to my searching key words. I also tried different key words like searching 'USB-C to USB-C' to ensure the accuracy instead of only 'UCB-C' which would return undesired results.

I did thie project on my own with no collaboration with others.

\section{Limitation}
We admit that only taking the number of reviews as the number of purchasement for different product may lead to some false positives and false negative results. Besides Walmart is not the most popular online shopping platform as Amazon and it could not accurately reflect the number of purchasement behaviors. Moreover, customers would not only decide to buy products online but also shop offline in stores. Despite all these limitations above, we believe our results and analysis above are still useful and insightful for better understanding the obsoleted chargers according to user behaviors.

\section{Future Work}
In recent years, advancements in wireless communication, secure data transfer, and AI-driven device optimization have reshaped the landscape of consumer electronics. From predictive energy management enabled by machine learning to secure connectivity protocols safeguarding data integrity, these innovations underline the interplay between technological progress and practical application.

Future advancements in charging technologies should further integrate innovations from these most advanced topics in like IoT \cite{wire1,wire3,10017581, 9523755,9340574,10.1145/3387514.3405861,9141221,9120764,10.1145/3356250.3360046,8737525,8694952,10.1145/3274783.3274846,10.1145/3210240.3210346,8486349,8117550,8057109,https://doi.org/10.1155/2017/5156164,10189210}, secure communication \cite{wire2, 10125074,285483,10.1145/3395351.3399367}, smart life \cite{MILLER2022100245,DBLP:journals/corr/abs-2112-15169,YAO2020100087,MILLER2020100089,8556650,10.1145/3127502.3127518,10.1145/3132479.3132480,GAO201718,10.1145/3769102.3770620}, machine learning \cite{10.1145/3460120.3484766,9709070,9444204,ning2021benchmarkingmachinelearningfast,8832180,8556807,8422243,chandrasekaran2022computervisionbasedparking,iqbal2021machinelearningartificialintelligence,pan2020endogenous} and so on. For instance, the development of wireless charging solutions informed by machine learning could enable more efficient energy transfer and minimize environmental impact. Predictive algorithms could optimize charging cycles based on user behavior and environmental conditions and reduce wear on batteries and extend their lifespan. Additionally, secure communication protocols can be explored to ensure safe and reliable data exchange in charging systems, particularly as devices become increasingly interconnected.

\section{Conclusion}
In conclusion, our exploration into the evolution of MacBook chargers, their environmental impact, and the economical considerations presents a nuanced view of technological progress. While innovation drives the development of more efficient, user-friendly, and potentially environmentally sustainable charging solutions, it also contributes to the growing issue of e-waste and demands significant investment from both manufacturers and consumers. Future studies should focus on the lifecycle analysis of emerging charging technologies to assess their true environmental footprint comprehensively. Attention should also be given to developing recycling and repurposing strategies that mitigate the environmental impact of obsolete chargers. Moreover, exploring the economic implications of adopting new technologies will provide valuable insights into the balance between cost, convenience, and sustainability. This multifaceted approach will help in making informed decisions that align technological advancement with environmental stewardship and economic viability, ensuring a sustainable future for electronic devices and their accessories.

\bibliographystyle{IEEEtran}
\bibliography{IEEEexample,zhu}

@article{IEEEexample:acharige2022review,
  title={Review of Electric Vehicle Charging Technologies, Configurations, and Architectures},
  author={Acharige, S. S. and Rajakaruna, S.},
  journal={Energies},
  volume={15},
  number={4},
  pages={1075},
  year={2022},
  publisher={Multidisciplinary Digital Publishing Institute}
}

@article{IEEEexample:overview_smartphones,
  title={Overview of Charging Technology Evolution in Smartphones},
  author={Alotaibi, F. and Alshehri, M.},
  journal={International Journal of Computer Applications},
  volume={182},
  number={45},
  pages={1--5},
  year={2019},
  publisher={Foundation of Computer Science (FCS)}
}

@article{IEEEexample:wireless_charging,
  title={Wireless Charging Technologies: Fundamentals, Standards, and Network Applications},
  author={Zhang, R. and Ho, C. K.},
  journal={IEEE Communications Surveys \& Tutorials},
  volume={17},
  number={2},
  pages={580--602},
  year={2015},
  publisher={IEEE}
}

@misc{IEEEexample:magsafe_is_back,
  title={Apple brings back the MagSafe connector — MagSafe 3},
  author={{Apple Inc.}},
  year={2021},
  howpublished={\url{https://www.apple.com/newsroom/2021/10/apple-brings-back-the-magsafe-connector/}}
}

@misc{IEEEexample:magsafe_charger_power,
  title={Apple’s New MagSafe Charger Packs Plenty of Power},
  author={{Apple Inc.}},
  year={2021},
  howpublished={\url{https://www.macrumors.com/2021/10/18/magsafe-charger-power/}}
}

@article{IEEEexample:moheb-alizadeh2023reverse,
  title={Reverse Logistics Network Design to Estimate the Economic and Environmental Impacts of Take-back Legislation: A Case Study for E-waste Management System in Washington State},
  author={Moheb-Alizadeh, H. and Freire, J.},
  journal={Resources, Conservation and Recycling},
  volume={175},
  pages={105926},
  year={2023},
  publisher={Elsevier}
}

@article{IEEEexample:ssd_embodied_carbon,
  title={The Dirty Secret of SSDs: Embodied Carbon},
  author={Koomey, J. and Berard, S.},
  journal={Environmental Research Letters},
  volume={14},
  number={3},
  pages={035003},
  year={2019},
  publisher={IOP Publishing}
}

@article{IEEEexample:ma2021technological,
  title={Technological Obsolescence},
  author={Ma, L. and Wang, S.},
  journal={Journal of Financial Economics},
  volume={140},
  number={1},
  pages={1--23},
  year={2021},
  publisher={Elsevier}
}

@incollection{IEEEexample:kowalkowski2017managing,
  title={Managing Technological Obsolescence in a Digitally Transformed SME},
  author={Kowalkowski, C. and Witell, L.},
  booktitle={Handbook of Research on Strategic Business Infrastructure Development and Contemporary Issues in Finance},
  pages={1--19},
  year={2017},
  publisher={IGI Global}
}

@misc{IEEEexample:global_ewaste_monitor_2024,
  title={The Global E-waste Monitor 2024},
  author={{United Nations University (UNU), International Telecommunication Union (ITU), \& International Solid Waste Association (ISWA)}},
  year={2024},
  howpublished={\url{https://globalewaste.org/}}
}

@article{IEEEexample:artificial_intelligence_sustainability,
  title={Artificial Intelligence for Sustainability: Facilitating Sustainable Smart Product-Service Systems with Computer Vision},
  author={Schreiber, M. and Meier, H.},
  journal={Procedia CIRP},
  volume={93},
  pages={3--8},
  year={2020},
  publisher={Elsevier}
}

@article{IEEEexample:sustainable_electronics_design,
  title={Sustainable Electronics Product Design and Manufacturing: State of Art Review},
  author={Khan, M. A. and Haque, M. E.},
  journal={Procedia Manufacturing},
  volume={35},
  pages={1285--1290},
  year={2019},
  publisher={Elsevier}
}

@article{YAO2020100087,
title = {Paris: Passive and continuous fetal heart monitoring system},
journal = {Smart Health},
volume = {17},
pages = {100087},
year = {2020},
issn = {2352-6483},
doi = {https://doi.org/10.1016/j.smhl.2019.100087},
url = {https://www.sciencedirect.com/science/article/pii/S2352648319300510},
author = {Yao Yao and Zeyu Ning and Qingquan Zhang and Ting Zhu}
}

@article{DBLP:journals/corr/abs-2112-15169,
  author       = {Sharlet Claros and
                  Wei Wang and
                  Ting Zhu},
  title        = {Investigations of Smart Health Reliability},
  journal      = {CoRR},
  volume       = {abs/2112.15169},
  year         = {2021},
  url          = {https://arxiv.org/abs/2112.15169},
  eprinttype    = {arXiv},
  eprint       = {2112.15169},
  bibsource    = {dblp computer science bibliography, https://dblp.org}
}

@article{MILLER2022100245,
title = {Radar-based monitoring system for medication tampering using data augmentation and multivariate time series classification},
journal = {Smart Health},
volume = {23},
pages = {100245},
year = {2022},
issn = {2352-6483},
doi = {https://doi.org/10.1016/j.smhl.2021.100245},
url = {https://www.sciencedirect.com/science/article/pii/S235264832100060X},
author = {Elishiah Miller and Zane MacFarlane and Seth Martin and Nilanjan Banerjee and Ting Zhu}
}

@inproceedings{wire1,
author = {Wang, Wei and Liu, Xin and Chi, Zicheng and Ray, Stuart and Zhu, Ting},
title = {Key Establishment for Secure Asymmetric Cross-Technology Communication},
year = {2024},
isbn = {9798400704826},
publisher = {Association for Computing Machinery},
address = {New York, NY, USA},
url = {https://doi-org.proxy.lib.ohio-state.edu/10.1145/3634737.3637670},
doi = {10.1145/3634737.3637670},
booktitle = {Proceedings of the 19th ACM Asia Conference on Computer and Communications Security},
pages = {412–422},
numpages = {11},
location = {Singapore, Singapore},
series = {ASIA CCS '24}
}

@inproceedings {wire2,
author = {Xin Liu and Wei Wang and Guanqun Song and Ting Zhu},
title = {{LightThief}: Your Optical Communication Information is Stolen behind the Wall},
booktitle = {32nd USENIX Security Symposium (USENIX Security 23)},
year = {2023},
isbn = {978-1-939133-37-3},
address = {Anaheim, CA},
pages = {5325--5339},
url = {https://www.usenix.org/conference/usenixsecurity23/presentation/liu-xin},
publisher = {USENIX Association},
month = aug
}

@misc{wire3,
      title={ML-based Secure Low-Power Communication in Adversarial Contexts}, 
      author={Guanqun Song and Ting Zhu},
      year={2022},
      eprint={2212.13689},
      archivePrefix={arXiv},
      primaryClass={cs.CR},
      url={https://arxiv.org/abs/2212.13689}, 
}

@INPROCEEDINGS{10017581,
  author={Wang, Wei and Chi, Zicheng and Liu, Xin and Bhaskar, Ananth Vishnu and Baingane, Ankit and Jahnige, Ryan and Zhang, Qingquan and Zhu, Ting},
  booktitle={MILCOM 2022 - 2022 IEEE Military Communications Conference (MILCOM)}, 
  title={A Secured Protocol for IoT Devices in Tactical Networks}, 
  year={2022},
  volume={},
  number={},
  pages={43-48},
  doi={10.1109/MILCOM55135.2022.10017581}}

@article{GAO201718,
title = {A smart medical system for dynamic closed-loop blood glucose-insulin control},
journal = {Smart Health},
volume = {1-2},
pages = {18-33},
year = {2017},
note = {Connected Health: Applications, Systems and Engineering Technologies (CHASE 2016)},
issn = {2352-6483},
doi = {https://doi.org/10.1016/j.smhl.2017.04.001},
url = {https://www.sciencedirect.com/science/article/pii/S2352648317300119},
author = {Jialin Gao and Ping Yi and Zicheng Chi and Ting Zhu}
}

@inproceedings{10.1145/3132479.3132480,
author = {Sui, Yu and Yi, Ping and Liu, Xin and Wang, Wei and Zhu, Ting},
title = {Optimization for charge station placement in electric vehicles energy network},
year = {2017},
isbn = {9781450355285},
publisher = {Association for Computing Machinery},
address = {New York, NY, USA},
url = {https://doi-org.proxy.lib.ohio-state.edu/10.1145/3132479.3132480},
doi = {10.1145/3132479.3132480},
booktitle = {Proceedings of the Workshop on Smart Internet of Things},
articleno = {1},
numpages = {6},
location = {San Jose, California},
series = {SmartIoT '17}
}

@inproceedings{10.1145/3127502.3127518,
author = {Liu, Xin and Wang, Wei and Zhu, Ting and Zhang, Qingquan and Yi, Ping},
title = {Poster: Smart Object-Oriented Dynamic Energy Management for Base Stations in Smart Cities},
year = {2017},
isbn = {9781450351416},
publisher = {Association for Computing Machinery},
address = {New York, NY, USA},
url = {https://doi-org.proxy.lib.ohio-state.edu/10.1145/3127502.3127518},
doi = {10.1145/3127502.3127518},
booktitle = {Proceedings of the 3rd Workshop on Experiences with the Design and Implementation of Smart Objects},
pages = {27–28},
numpages = {2},
location = {Snowbird, Utah, USA},
series = {SMARTOBJECTS '17}
}

@INPROCEEDINGS{8556650,
  author={Hu, Yunlong and Yi, Ping and Sui, Yu and Zhang, Zongrui and Yao, Yao and Wang, Wei and Zhu, Ting},
  booktitle={2018 Third International Conference on Security of Smart Cities, Industrial Control System and Communications (SSIC)}, 
  title={Dispatching and Distributing Energy in Energy Internet under Energy Dilemma}, 
  year={2018},
  volume={},
  number={},
  pages={1-5},
  doi={10.1109/SSIC.2018.8556650}}

@article{MILLER2020100089,
title = {RadSense: Enabling one hand and no hands interaction for sterile manipulation of medical images using Doppler radar},
journal = {Smart Health},
volume = {15},
pages = {100089},
year = {2020},
issn = {2352-6483},
doi = {https://doi.org/10.1016/j.smhl.2019.100089},
url = {https://www.sciencedirect.com/science/article/pii/S2352648319300534},
author = {Elishiah Miller and Zheng Li and Helena Mentis and Adrian Park and Ting Zhu and Nilanjan Banerjee}
}

@inproceedings{10.1145/3460120.3484766,
author = {Wang, Wei and Yao, Yao and Liu, Xin and Li, Xiang and Hao, Pei and Zhu, Ting},
title = {I Can See the Light: Attacks on Autonomous Vehicles Using Invisible Lights},
year = {2021},
isbn = {9781450384544},
publisher = {Association for Computing Machinery},
address = {New York, NY, USA},
url = {https://doi-org.proxy.lib.ohio-state.edu/10.1145/3460120.3484766},
doi = {10.1145/3460120.3484766},
booktitle = {Proceedings of the 2021 ACM SIGSAC Conference on Computer and Communications Security},
pages = {1930–1944},
numpages = {15},
location = {Virtual Event, Republic of Korea},
series = {CCS '21}
}

@ARTICLE{9523755,
  author={Cheng, Long and Kong, Linghe and Gu, Yu and Niu, Jianwei and Zhu, Ting and Liu, Cong and Mumtaz, Shahid and He, Tian},
  journal={IEEE Transactions on Wireless Communications}, 
  title={Collision-Free Dynamic Convergecast in Low-Duty-Cycle Wireless Sensor Networks}, 
  year={2022},
  volume={21},
  number={3},
  pages={1665-1680},
  doi={10.1109/TWC.2021.3105983}}

@ARTICLE{9340574,
  author={Chi, Zicheng and Li, Yan and Sun, Hongyu and Huang, Zhichuan and Zhu, Ting},
  journal={IEEE/ACM Transactions on Networking}, 
  title={Simultaneous Bi-Directional Communications and Data Forwarding Using a Single ZigBee Data Stream}, 
  year={2021},
  volume={29},
  number={2},
  pages={821-833},
  doi={10.1109/TNET.2021.3054339}}

@inproceedings{10.1145/3387514.3405861,
author = {Chi, Zicheng and Liu, Xin and Wang, Wei and Yao, Yao and Zhu, Ting},
title = {Leveraging Ambient LTE Traffic for Ubiquitous Passive Communication},
year = {2020},
isbn = {9781450379557},
publisher = {Association for Computing Machinery},
address = {New York, NY, USA},
url = {https://doi-org.proxy.lib.ohio-state.edu/10.1145/3387514.3405861},
doi = {10.1145/3387514.3405861},
booktitle = {Proceedings of the Annual Conference of the ACM Special Interest Group on Data Communication on the Applications, Technologies, Architectures, and Protocols for Computer Communication},
pages = {172–185},
numpages = {14},
location = {Virtual Event, USA},
series = {SIGCOMM '20}
}

@inproceedings{10.1145/3769102.3770620,
author = {Song, Guanqun and Li, Yan and Zhu, Ting},
title = {A Metal Sensing and Biometric-based Tracking System},
year = {2025},
isbn = {9798400722387},
publisher = {Association for Computing Machinery},
address = {New York, NY, USA},
url = {https://doi-org.proxy.lib.ohio-state.edu/10.1145/3769102.3770620},
doi = {10.1145/3769102.3770620},
booktitle = {Proceedings of the Tenth ACM/IEEE Symposium on Edge Computing},
articleno = {23},
numpages = {17},
location = {the Hilton Arlington National Landing, Arlington, VA, USA},
series = {SEC '25}
}

@ARTICLE{9141221,
  author={Pan, Yan and Li, Shining and Li, Bingqi and Zhang, Yu and Yang, Zhe and Guo, Bin and Zhu, Ting},
  journal={IEEE Communications Magazine}, 
  title={CDD: Coordinating Data Dissemination in Heterogeneous IoT Networks}, 
  year={2020},
  volume={58},
  number={6},
  pages={84-89},
  doi={10.1109/MCOM.001.1900473}}

@INPROCEEDINGS{9120764,
  author={Tao, Yinrong and Xiao, Sheng and Hao, Bin and Zhang, Qingquan and Zhu, Ting and Chen, Zhuo},
  booktitle={2020 IEEE Wireless Communications and Networking Conference (WCNC)}, 
  title={WiRE: Security Bootstrapping for Wireless Device-to-Device Communication}, 
  year={2020},
  volume={},
  number={},
  pages={1-7},
  doi={10.1109/WCNC45663.2020.9120764}}

@inproceedings{10.1145/3356250.3360046,
author = {Chi, Zicheng and Li, Yan and Liu, Xin and Yao, Yao and Zhang, Yanchao and Zhu, Ting},
title = {Parallel inclusive communication for connecting heterogeneous IoT devices at the edge},
year = {2019},
isbn = {9781450369503},
publisher = {Association for Computing Machinery},
address = {New York, NY, USA},
url = {https://doi-org.proxy.lib.ohio-state.edu/10.1145/3356250.3360046},
doi = {10.1145/3356250.3360046},
booktitle = {Proceedings of the 17th Conference on Embedded Networked Sensor Systems},
pages = {205–218},
numpages = {14},
location = {New York, New York},
series = {SenSys '19}
}

@INPROCEEDINGS{8737525,
  author={Wang, Wei and Liu, Xin and Yao, Yao and Pan, Yan and Chi, Zicheng and Zhu, Ting},
  booktitle={IEEE INFOCOM 2019 - IEEE Conference on Computer Communications}, 
  title={CRF: Coexistent Routing and Flooding using WiFi Packets in Heterogeneous IoT Networks}, 
  year={2019},
  volume={},
  number={},
  pages={19-27},
  doi={10.1109/INFOCOM.2019.8737525}}

@ARTICLE{8694952,
  author={Chi, Zicheng and Li, Yan and Sun, Hongyu and Yao, Yao and Zhu, Ting},
  journal={IEEE/ACM Transactions on Networking}, 
  title={Concurrent Cross-Technology Communication Among Heterogeneous IoT Devices}, 
  year={2019},
  volume={27},
  number={3},
  pages={932-947},
  doi={10.1109/TNET.2019.2908754}}

@inproceedings{10.1145/3274783.3274846,
author = {Li, Yan and Chi, Zicheng and Liu, Xin and Zhu, Ting},
title = {Passive-ZigBee: Enabling ZigBee Communication in IoT Networks with 1000X+ Less Power Consumption},
year = {2018},
isbn = {9781450359528},
publisher = {Association for Computing Machinery},
address = {New York, NY, USA},
url = {https://doi-org.proxy.lib.ohio-state.edu/10.1145/3274783.3274846},
doi = {10.1145/3274783.3274846},
booktitle = {Proceedings of the 16th ACM Conference on Embedded Networked Sensor Systems},
pages = {159–171},
numpages = {13},
location = {Shenzhen, China},
series = {SenSys '18}
}

@inproceedings{10.1145/3210240.3210346,
author = {Li, Yan and Chi, Zicheng and Liu, Xin and Zhu, Ting},
title = {Chiron: Concurrent High Throughput Communication for IoT Devices},
year = {2018},
isbn = {9781450357203},
publisher = {Association for Computing Machinery},
address = {New York, NY, USA},
url = {https://doi-org.proxy.lib.ohio-state.edu/10.1145/3210240.3210346},
doi = {10.1145/3210240.3210346},
booktitle = {Proceedings of the 16th Annual International Conference on Mobile Systems, Applications, and Services},
pages = {204–216},
numpages = {13},
location = {Munich, Germany},
series = {MobiSys '18}
}

@INPROCEEDINGS{8486349,
  author={Wang, Wei and Xie, Tiantian and Liu, Xin and Zhu, Ting},
  booktitle={IEEE INFOCOM 2018 - IEEE Conference on Computer Communications}, 
  title={ECT: Exploiting Cross-Technology Concurrent Transmission for Reducing Packet Delivery Delay in IoT Networks}, 
  year={2018},
  volume={},
  number={},
  pages={369-377},
  doi={10.1109/INFOCOM.2018.8486349}}

@INPROCEEDINGS{8117550,
  author={Chi, Zicheng and Li, Yan and Yao, Yao and Zhu, Ting},
  booktitle={2017 IEEE 25th International Conference on Network Protocols (ICNP)}, 
  title={PMC: Parallel multi-protocol communication to heterogeneous IoT radios within a single WiFi channel}, 
  year={2017},
  volume={},
  number={},
  pages={1-10},
  doi={10.1109/ICNP.2017.8117550}}

@INPROCEEDINGS{8057109,
  author={Chi, Zicheng and Huang, Zhichuan and Yao, Yao and Xie, Tiantian and Sun, Hongyu and Zhu, Ting},
  booktitle={IEEE INFOCOM 2017 - IEEE Conference on Computer Communications}, 
  title={EMF: Embedding multiple flows of information in existing traffic for concurrent communication among heterogeneous IoT devices}, 
  year={2017},
  volume={},
  number={},
  pages={1-9},
  doi={10.1109/INFOCOM.2017.8057109}}

@article{https://doi.org/10.1155/2017/5156164,
author = {Sun, Hongyu and Fang, Zhiyi and Liu, Qun and Lu, Zheng and Zhu, Ting},
title = {Enabling LTE and WiFi Coexisting in 5 GHz for Efficient Spectrum Utilization},
journal = {Journal of Computer Networks and Communications},
volume = {2017},
number = {1},
pages = {5156164},
doi = {https://doi.org/10.1155/2017/5156164},
url = {https://onlinelibrary.wiley.com/doi/abs/10.1155/2017/5156164},
eprint = {https://onlinelibrary.wiley.com/doi/pdf/10.1155/2017/5156164},
year = {2017}
}

@INPROCEEDINGS{9709070,
  author={Wang, Wei and Ning, Zeyu and Iradukunda, Hugues and Zhang, Qingquan and Zhu, Ting and Yi, Ping},
  booktitle={2021 IEEE/ACM Symposium on Edge Computing (SEC)}, 
  title={MailLeak: Obfuscation-Robust Character Extraction Using Transfer Learning}, 
  year={2021},
  volume={},
  number={},
  pages={459-464},
  doi={10.1145/3453142.3491421}}

@ARTICLE{9444204,
  author={Han, Dianqi and Li, Ang and Zhang, Lili and Zhang, Yan and Li, Jiawei and Li, Tao and Zhu, Ting and Zhang, Yanchao},
  journal={IEEE/ACM Transactions on Networking}, 
  title={Deep Learning-Guided Jamming for Cross-Technology Wireless Networks: Attack and Defense}, 
  year={2021},
  volume={29},
  number={5},
  pages={1922-1932},
  doi={10.1109/TNET.2021.3082839}}

@misc{ning2021benchmarkingmachinelearningfast,
      title={Benchmarking Machine Learning: How Fast Can Your Algorithms Go?}, 
      author={Zeyu Ning and Hugues Nelson Iradukunda and Qingquan Zhang and Ting Zhu},
      year={2021},
      eprint={2101.03219},
      archivePrefix={arXiv},
      primaryClass={cs.LG},
      url={https://arxiv.org/abs/2101.03219}, 
}

@ARTICLE{8832180,
  author={Xia, Zhiyang and Yi, Ping and Liu, Yunyu and Jiang, Bo and Wang, Wei and Zhu, Ting},
  journal={IEEE Transactions on Multimedia}, 
  title={GENPass: A Multi-Source Deep Learning Model for Password Guessing}, 
  year={2020},
  volume={22},
  number={5},
  pages={1323-1332},
  doi={10.1109/TMM.2019.2940877}}

@INPROCEEDINGS{8556807,
  author={Meng, Yishuang and Yi, Ping and Guo, Xuejun and Gu, Wen and Liu, Xin and Wang, Wei and Zhu, Ting},
  booktitle={2018 Third International Conference on Security of Smart Cities, Industrial Control System and Communications (SSIC)}, 
  title={Detection for Pulmonary Nodules using RGB Channel Superposition Method in Deep Learning Framework}, 
  year={2018},
  volume={},
  number={},
  pages={1-8},
  doi={10.1109/SSIC.2018.8556807}}

@INPROCEEDINGS{8422243,
  author={Liu, Yunyu and Xia, Zhiyang and Yi, Ping and Yao, Yao and Xie, Tiantian and Wang, Wei and Zhu, Ting},
  booktitle={2018 IEEE International Conference on Communications (ICC)}, 
  title={GENPass: A General Deep Learning Model for Password Guessing with PCFG Rules and Adversarial Generation}, 
  year={2018},
  volume={},
  number={},
  pages={1-6},
  doi={10.1109/ICC.2018.8422243}}

@ARTICLE{10189210,
  author={Liu, Xin and Chi, Zicheng and Wang, Wei and Yao, Yao and Hao, Pei and Zhu, Ting},
  journal={IEEE/ACM Transactions on Networking}, 
  title={High-Granularity Modulation for OFDM Backscatter}, 
  year={2024},
  volume={32},
  number={1},
  pages={338-351},
  doi={10.1109/TNET.2023.3286880}}

@ARTICLE{10125074,
  author={Yao, Yao and Li, Yan and Zhu, Ting},
  journal={IEEE Transactions on Mobile Computing}, 
  title={Interference-Negligible Privacy-Preserved Shield for RF Sensing}, 
  year={2024},
  volume={23},
  number={5},
  pages={3576-3588},
  doi={10.1109/TMC.2023.3276930}}

@inproceedings {285483,
author = {Ang Li and Jiawei Li and Dianqi Han and Yan Zhang and Tao Li and Ting Zhu and Yanchao Zhang},
title = {{PhyAuth}: {Physical-Layer} Message Authentication for {ZigBee} Networks},
booktitle = {32nd USENIX Security Symposium (USENIX Security 23)},
year = {2023},
isbn = {978-1-939133-37-3},
address = {Anaheim, CA},
pages = {1--18},
url = {https://www.usenix.org/conference/usenixsecurity23/presentation/li-ang},
publisher = {USENIX Association},
month = aug
}

@misc{chandrasekaran2022computervisionbasedparking,
      title={Computer Vision Based Parking Optimization System}, 
      author={Siddharth Chandrasekaran and Jeffrey Matthew Reginald and Wei Wang and Ting Zhu},
      year={2022},
      eprint={2201.00095},
      archivePrefix={arXiv},
      primaryClass={cs.CV},
      url={https://arxiv.org/abs/2201.00095}, 
}

@misc{iqbal2021machinelearningartificialintelligence,
      title={Machine Learning and Artificial Intelligence in Next-Generation Wireless Network}, 
      author={Wafeeq Iqbal and Wei Wang and Ting Zhu},
      year={2021},
      eprint={2202.01690},
      archivePrefix={arXiv},
      primaryClass={cs.NI},
      url={https://arxiv.org/abs/2202.01690}, 
}

@inproceedings{10.1145/3395351.3399367,
author = {Chi, Zicheng and Li, Yan and Liu, Xin and Wang, Wei and Yao, Yao and Zhu, Ting and Zhang, Yanchao},
title = {Countering cross-technology jamming attack},
year = {2020},
isbn = {9781450380065},
publisher = {Association for Computing Machinery},
address = {New York, NY, USA},
url = {https://doi-org.proxy.lib.ohio-state.edu/10.1145/3395351.3399367},
doi = {10.1145/3395351.3399367},
booktitle = {Proceedings of the 13th ACM Conference on Security and Privacy in Wireless and Mobile Networks},
pages = {99–110},
numpages = {12},
location = {Linz, Austria},
series = {WiSec '20}
}

@article{pan2020endogenous,
  title={Endogenous Security Defense against Deductive Attack: When Artificial Intelligence Meets Active Defense for Online Service},
  author={Pan, Yan and Li, Shining and Li, Bingqi and Zhang, Yu and Yang, Zhe and Guo, Bin and Zhu, Ting},
  journal={IEEE Communications Magazine},
  volume={58},
  number={6},
  pages={84--89},
  year={2020},
  publisher={IEEE-INST ELECTRICAL ELECTRONICS ENGINEERS INC 445 HOES LANE, PISCATAWAY, NJ~…}
}

\end{document}